\documentclass[twocolumn,showpacs,amsmath,amssymb,superscriptaddress,prx]{revtex4-1}

\usepackage{bm}
\usepackage{longtable}

\usepackage{epsfig}
\usepackage{graphicx}
\usepackage{subfigure}
\usepackage{comment}
\usepackage{amsfonts}
\usepackage{amsmath}
\usepackage{amssymb}
\usepackage{subfigure}

\usepackage{color}

\usepackage[colorlinks,bookmarks=false,citecolor=blue,linkcolor=blue,urlcolor=blue]{hyperref}

\newcommand{\bs}{\boldsymbol}
\newcommand{\be}{\begin{equation}}
\newcommand{\ee}{\end{equation}}
\newcommand{\bea}{\begin{eqnarray}}
\newcommand{\eea}{\end{eqnarray}}

\def\bs{\boldsymbol}
\def\vec{\mathbf}
\def\mc{\mathcal}

\usepackage{times}

\begin{document}

\title{ Mechanism of basal-plane antiferromagnetism in the spin-orbit driven iridate Ba$_2$IrO$_4$}

\author{Vamshi M.~Katukuri}
\affiliation {Institute for Theoretical Solid State Physics, IFW Dresden, Helmholtzstr.~20, 01069 Dresden, Germany}

\author{Viktor Yushankhai}
\affiliation{Institute for Theoretical Solid State Physics, IFW Dresden, Helmholtzstr.~20, 01069 Dresden, Germany}
\affiliation{Joint Institute for Nuclear Research, Joliot-Curie 6, 141980 Dubna, Russia}

\author{Liudmila Siurakshina}
\affiliation{Joint Institute for Nuclear Research, Joliot-Curie 6, 141980 Dubna, Russia}
\affiliation{Max-Planck-Institut f\"{u}r Physik komplexer Systeme, N\"{o}thnitzer Str.~38, 01187 Dresden, Germany}

\author{Jeroen van den Brink}
\affiliation{Institute for Theoretical Solid State Physics, IFW Dresden, Helmholtzstr.~20, 01069 Dresden, Germany}

\author{Liviu Hozoi}
\affiliation {Institute for Theoretical Solid State Physics, IFW Dresden, Helmholtzstr.~20, 01069 Dresden, Germany}

\author{Ioannis Rousochatzakis}
\affiliation {Institute for Theoretical Solid State Physics, IFW Dresden, Helmholtzstr.~20, 01069 Dresden, Germany}
\affiliation{Max-Planck-Institut f\"{u}r Physik komplexer Systeme, N\"{o}thnitzer Str.~38, 01187 Dresden, Germany}

\begin{abstract}
By {\it ab initio} many-body quantum chemistry calculations, we determine the strength of the
symmetric anisotropy in the $5d^5$ $j\!\approx\!1/2$ layered material Ba$_2$IrO$_4$.
While the calculated anisotropic couplings come out in the range of a few meV,
orders of magnitude stronger than in analogous $3d$ transition-metal compounds,
the Heisenberg superexchange still defines the largest energy scale.
The {\it ab initio} results reveal that individual layers of Ba$_2$IrO$_4$ provide a close 
realization of the quantum spin-1/2 Heisenberg-compass model on the square lattice.
We show that the experimentally observed basal-plane antiferromagnetism can be accounted for by
including additional interlayer interactions and the associated order-by-disorder quantum-mechanical
effects, in analogy to undoped layered cuprates.
\end{abstract}

\date\today


\maketitle

\section{Introduction}\label{sec:I}
The few varieties of square-lattice effective spin models are emblematic in modern quantum magnetism
and extensively investigated in relation to layered superconducting materials such as the copper
oxides \cite{Bednorz_Muller_HTc} and the iron pnictides/chalcogenides \cite{FeAs_Superconductor_discovery}.
While the dominant magnetic energy scale is set in these systems by the isotropic Heisenberg exchange
between nearest-neighbor (NN)~\cite{CuO_PALee_rev_06} and possibly next-NN sites~\cite{FeAs_magnetism_PRL,*FeAs_yildirim_rev_09},
there are many examples where the smaller, anisotropic terms become important too, e.g., for correctly
describing the antiferromagnetic (AF) ordering pattern in La$_2$CuO$_4$ \cite{Keimer_anisotropy_cuprates}
or in the cuprate oxychlorides \cite{SrCuOCl_afm_PRB,SrCuOCl_INS_ZPB}. 
This topic, the role of anisotropic interactions in transition-metal compounds has lately received a
new impetus with recent insights into the basic electronic structure of $5d$ systems such as the $5d^5$
iridium oxides.
Here, a subtle interplay between spin-orbit interactions and sizable electron correlations gives rise
to insulating ground states and well protected magnetic moments
\cite{214Ir_kim_2009,IrO_mott_kim_08,IrO_kitaev_jackeli_09,Kim12,BoseggiaPRL_13,Honeycomb_NaIrO_Choi_2012}.
Due to the strong spin-orbit couplings, however, these magnetic moments are best described as
effective $j\!\approx\!1/2$ entities \cite{SOC_d5_thornley68,IrO_mott_kim_08,IrO_kitaev_jackeli_09}
and the effective anisotropic exchange parameters are orders of magnitude larger than in $3d$
transition-metal compounds.
For the square-lattice system Sr$_2$IrO$_4$, for instance, Dzyaloshinskii-Moriya (DM) interactions as large
as one quarter of the NN AF superexchange have been predicted \cite{Ir214_BHKim_2012,Ir214_perkins_13}
while in honeycomb iridates the symmetric Kitaev exchange is believed to be even larger than the Heisenberg
interaction \cite{ZigZag_KH_chalopka_12,Mazin_Na213_2013,Ir213_katukuri_13,Na2IrO3_KH_Hlynur_2013}.

Valuable insights into the role of different superexchange processes in correlated $d$-metal
oxides come from the detailed analysis of extended multiorbital Hubbard-type models.
The foundations of superexchange theory were laid as early as the 50's with the work of Anderson,
Goodenough, and Kanamori \cite{Anderson_1950,*Kanamori01021957_1,*Kanamori01021957_2,*Goodenough1958287,*Kanamori195987}.
Standard approaches within this theoretical framework proved to be extremely useful in, e.g., better
understanding the origin and relative strength of the anisotropic couplings in layered cuprates
\cite{Yildirim96,Aharony98}.
In two-dimensional (2D) iridates, on the other hand, much less information is presently available on
the magnitude of various electronic-structure parameters that enter the superexchange models.
While estimates for these effective electronic-structure parameters are normally based on either
density-functional band-structure calculations \cite{Martins_PRL_2011_Ir_Rh,IrO_mott_arita_12,Mazin_Na213_2013,Ir214_BHKim_2012,Ir214_perkins_13}
or experiments \cite{IrO_kitaev_jackeli_09,ZigZag_KH_chalopka_12,Kim12,Na2IrO3_KH_Hlynur_2013,Honeycomb_NaIrO_Choi_2012},
we here rely on many-body quantum chemistry methods to directly obtain an {\it ab initio}
assessment of both the NN Heisenberg exchange and the anisotropic couplings on the square
lattice of Ba$_2$IrO$_4$.
%
%
%
%
Our study reveals uniaxial symmetric anisotropy that is bond dependent, thus giving rise to quantum
compass interaction terms~\cite{KK1973} superimposed onto the much stronger (due to the 180$^{\circ}$ bond
geometry) isotropic Heisenberg exchange.
We also show that the resulting Heisenberg-compass model for individual layers of Ba$_2$IrO$_4$ is not
sufficient to explain the AF ground-state ordering pattern
inferred from recent resonant magnetic scattering measurements, with spins ordered along the [110]
direction \cite{BoseggiaPRL_13}.
To rationalize the latter, we carry out a detailed analysis of the role of interlayer couplings
and the associated order-by-disorder phenomena.
An extended three-dimensional (3D) spin Hamiltonian based on NN exchange terms as found in the
{\em ab initio} quantum chemistry calculations and additional farther-neighbor interlayer
exchange integrals turns out to provide a realistic starting point to explain the magnetism of
Ba$_2$IrO$_4$.

\begin{figure}[!b]
\includegraphics[angle=0,width=8.0cm]{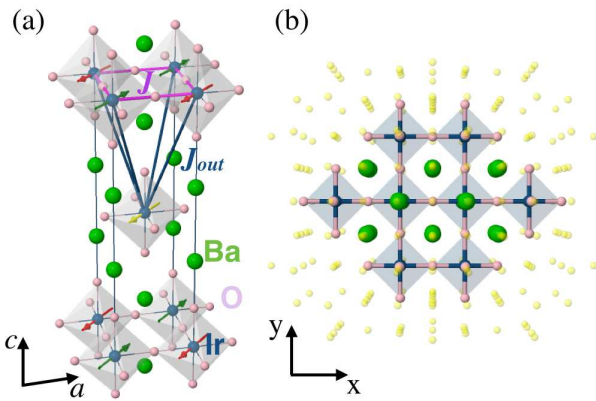}
\caption{a) Layered crystal structure of Ba$_2$IrO$_4$.
The in-plane and interlayer exchange paths are shown.
b) Sketch of the cluster used for the calculation of the magnetic interactions, see text.
Ir, O, and Ba ions are shown in blue, pink, and green, respectively.}
\label{figBa214}
\end{figure}


\section{General considerations}\label{sec:II}

The magnetically active sites, the Ir$^{4+}$ ions, have a $5d^5$ valence electron configuration in
Ba$_2$IrO$_4$, which under strong octahedral crystal-field and spin-orbit interactions yields an
effective $j\!\approx\!1/2$ Kramers-doublet ground state, see Refs.~\cite{SOC_d5_thornley68,IrO_kitaev_jackeli_09}
and \cite{BoseggiaPRL_13,Ba214_rixs_moretti_2014}.
The exchange interactions between such pseudospin entities involve both isotropic Heisenberg and anisotropic
terms.
For a pair of NN pseudospins $\tilde{\vec{S}}_i$ and $\tilde{\vec{S}}_j$, the most general bilinear spin
Hamiltonian can be cast in the form 
\be\label{a1}
\mc{H}_{ij} = J_{ij}\, \tilde{\vec{S}}_i\!\cdot\!\tilde{\vec{S}}_j 
+{\bf D}_{ij}\!\cdot\! {\tilde {\bf S}}_i\!\times\! {\tilde {\bf S}}_j 
+\tilde{\vec{S}}_i\!\cdot\!\bs{\Gamma}_{ij}\!\cdot\!\tilde{\vec{S}}_j\,,
\ee
where $J_{ij}$ is the isotropic Heisenberg exchange, the vector ${\bf D}_{ij}$ defines the DM anisotropy,
and $\bs{\Gamma}_{ij}$ is a symmetric traceless second-rank tensor that describes the symmetric portion
of the exchange anisotropy. 
Depending on various geometrical details and the choice of the reference frame, some elements
of the DM vector and/or of the $\Gamma^{\alpha\beta}_{ij}$ tensor may be zero.
For the square lattice of corner-sharing IrO$_6$ octahedra in Ba$_2$IrO$_4$, the symmetry of each
block of two NN octahedra is $D_{2h}$, with inversion symmetry at the bridging oxygen site
\cite{Ba214_structure}. Given the inversion center, the DM anisotropy vanishes.
The remaining symmetries require that in the $\{xyz\}$ frame, with $x$ along the Ir-Ir link and $z$
orthogonal to the IrO$_2$ layers, $\bs{\Gamma}_{ij}$ is diagonal.
The two-site effective spin Hamiltonian for an Ir-Ir link along the $x$ axis can then be written as
\be\label{a2}
\mc{H}_{\langle ij\rangle \parallel x} = J\, \tilde{\vec{S}}_i\!\cdot\!\tilde{\vec{S}}_j  
+\Gamma_{\parallel}  \tilde{S}_i^{x} \tilde{S}_j^{x} 
+\Gamma_{\perp}  \tilde{S}_i^{y} \tilde{S}_j^{y} 
+\Gamma_{zz} \tilde{S}_i^{z} \tilde{S}_j^{z}\,,
\ee
with $\Gamma_{zz}= -(\Gamma_{\parallel}+\Gamma_{\perp})$ since $\bs{\Gamma}$ is traceless.
Due to the four-fold $z$-axis symmetry, we analogously have
\be\label{a3}
\mc{H}_{\langle ij\rangle \parallel y} = J\, \tilde{\vec{S}}_i\!\cdot\!\tilde{\vec{S}}_j  +
                    \Gamma_{\parallel}  \tilde{S}_i^{y} \tilde{S}_j^{y} +
                     \Gamma_{\perp}  \tilde{S}_i^{x} \tilde{S}_j^{x} +
                     \Gamma_{zz}  \tilde{S}_i^{z} \tilde{S}_j^{z}
\ee
for bonds along the $y$ axis.
The eigenstates of (\ref{a2}) are the singlet
$|\Psi_{\mathrm{S}}\rangle\!=\!\frac{|\uparrow\downarrow\rangle - |\downarrow\uparrow\rangle}{\sqrt2}$
and the three ``triplet'' components 
$|\Psi_{\mathrm{1}}\rangle\!=\!\frac{|\uparrow\downarrow\rangle + |\downarrow\uparrow\rangle}{\sqrt2}$,
$|\Psi_{\mathrm{2}}\rangle\!=\!\frac{|\uparrow\uparrow\rangle + |\downarrow\downarrow\rangle}{\sqrt2}$,
$|\Psi_{\mathrm{3}}\rangle\!=\!\frac{|\uparrow\uparrow\rangle - |\downarrow\downarrow\rangle}{\sqrt2}$.
The corresponding eigenvalues are
\bea\label{a4}
E_{\mathrm{S}}= -\frac{3}{4}J, \
E_{1}=\frac{1}{4}J + \frac{1}{2}\left(\Gamma_{\parallel}+\Gamma_{\perp}\right),\nonumber \\
E_{2}=\frac{1}{4}J - \frac{1}{2}\Gamma_{\perp}, \ \ \
E_{3}=\frac{1}{4}J - \frac{1}{2}\Gamma_{\parallel}\,.
\eea
For $D_{2h}$ symmetry of the two-octahedra unit, the four low-lying (spin-orbit) states,
$|\Psi_{\mathrm{S}}\rangle$, $|\Psi_{\mathrm{1}}\rangle$, $|\Psi_{\mathrm{2}}\rangle$, and $|\Psi_{\mathrm{3}}\rangle$,
transform according to the $A_{1g}$, $B_{2u}$, $B_{1u}$, and $A_{1u}$ irreducible representations,
respectively \cite{Ir213_katukuri_13}.
As discussed in the following, this symmetry analysis is useful in determining the nature of each of
the low-lying many-body states in the quantum chemistry calculations.

\section{Quantum chemistry calculations}\label{sec:III}

\subsection{Computational details}

The effective magnetic coupling constants are obtained on the basis of multireference
configuration-interaction (MRCI) calculations\cite{book_QC_00} on units of two corner-sharing
IrO$_6$ octahedra.
Since it is important to accurately describe the charge distribution at sites in the immediate
neighborhood \cite{qc_NNs_degraaf_99,CuO2_dd_hozoi11,CuO_ZRB_hozoi_07}, we also include in the
actual cluster the closest 16 Ba ions and the six adjacent IrO$_6$ octahedra around the reference
[Ir$_2$O$_{11}$] fragment, see Fig.~\ref{figBa214} and also Refs.~\onlinecite{Ir213_katukuri_13,214Ir_vmk_2012,113Ir_bogdanov_2012,*CaIrO3_moretti_14,Os227_bogdanov_12}.
To make the whole analysis tractable, we however replaced the six Ir$^{4+}$ $d^5$ NN's by
closed-shell Pt$^{4+}$ $d^6$ ions, a usual procedure in quantum chemistry investigations on
$d$-metal systems \cite{Ir213_katukuri_13,214Ir_vmk_2012,113Ir_bogdanov_2012,*CaIrO3_moretti_14,Os227_bogdanov_12,Na2V2O5_hozoi_02,SIA_Fe_maurice_2013}.
The extended solid-state surroundings were modeled as a large array of point charges fitted
to reproduce the crystal Madelung field in the cluster region.
We used the crystal structure reported by Okabe {\it et al.}\cite{Ba214_structure}

All calculations were performed with the {\sc molpro} quantum chemistry software
\cite{molpro_brief}.
Energy-consistent relativistic pseudopotentials from the standard {\sc molpro} library were used
for Ir \cite{ECP_Stoll_2} and Ba \cite{FuentealbaJPB}.
The valence orbitals at the central Ir sites were described by basis sets of quadruple-zeta
quality supplemented with two $f$ polarization functions \cite{ECP_Stoll_2} while for the ligand bridging the two
magnetically active Ir ions we applied quintuple-zeta valence basis sets and four $d$ polarization
functions \cite{GBas_molpro_2p}.
The other O's at the two central octahedra were modeled by triple-zeta valence basis sets
\cite{GBas_molpro_2p}.
For the additional ligands coordinating the six adjacent $5d$ sites we used minimal atomic-natural-orbital
basis functions \cite{ANOs_pierloot_95}.
At those adjacent $5d$ sites we applied triple-zeta valence basis sets~\cite{ECP_Stoll_2}.

Multiconfiguration reference wave functions were first generated by complete-active-space
self-consistent-field (CASSCF) calculations \cite{book_QC_00}.
The active space is here given by five electrons and three ($t_{2g}$) orbitals at each of the
two magnetically active Ir sites.
The orbitals were optimized for an average of the lowest nine singlet and the nine triplet states
arising from such an active space.
All these states entered the spin-orbit calculations, both at the CASSCF and MRCI levels.
In the MRCI treatment, single and double excitations from the six Ir $t_{2g}$ orbitals
and the $2p$ shell of the bridging ligand site are taken into account.
Similar strategies of explicitly dealing only with selected groups of localized ligand orbitals
were adopted in earlier studies on both $3d$ \cite{QC_J_fink94,NOCI_J_oosten96,NOCI_J_hozoi03,J_ligand_calzado03}
and $5d$~\cite{Ir213_katukuri_13,214Ir_vmk_2012,113Ir_bogdanov_2012,*CaIrO3_moretti_14,Os227_bogdanov_12} compounds, with results
in good agreement with the experiment
\cite{214Ir_vmk_2012,113Ir_bogdanov_2012,*CaIrO3_moretti_14,NOCI_J_oosten96,NOCI_J_hozoi03,J_ligand_calzado03}.
To separate the metal $5d$ and O $2p$ valence orbitals into different groups, we used the
orbital localization module available in {\sc molpro}.
The MRCI was performed for each spin multiplicity, singlet or triplet, as a nine-root
calculation.

To obtain information on the magnitude of the direct exchange, we additionally carried out
single-configuration restricted open-shell Hartree-Fock (ROHF) calculations~\cite{book_QC_00}.
The latter were performed as frozen-orbital calculations, i.e., we used the orbitals
obtained by CASSCF (see above), without further optimization.

The spin-orbit treatment was carried out according to the procedure described in Ref.~[\onlinecite{SOC_molpro}].
To determine the nature of each spin-orbit state we explicitly compute with {\sc molpro} the
dipole and quadrupole transition matrix elements among those four low-lying states describing
the magnetic spectrum of two corner-sharing octahedra, see Table\,I and the next subsection.
Standard selection rules and the nonzero dipole and quadrupole matrix elements in the quantum
chemistry outputs then clearly indicate which state is which, see also the analysis and
discussion in Ref.~[\onlinecite{Ir213_katukuri_13}].

\subsection{{\it Ab initio} results}

Of the 36 spin-orbit states that are obtained in the {\it ab initio} calculations,
the low-lying four are listed in Table\,\ref{table_Js}
\footnote{
The higher-lying spin-orbit states imply an excitation energy of at least 0.6 eV.
This gap concerns the $j\!=\!1/2$ to $j\!=\!3/2$ transitions \cite{Kim12,RhIr214_vmk_13}
}.
These four states are further mapped onto the eigenvalues of the effective spin Hamiltonian in (\ref{a2}).
Energy splittings and the associated effective magnetic couplings are provided at three levels of approximation:
single-configuration ROHF (HF+SOC), CASSCF (CAS+SOC), and MRCI (CI+SOC). 
It is seen that at all levels of theory two of the triplet components, $\Psi_{\mathrm{1}}$ and
$\Psi_{\mathrm{2}}$, are degenerate
\footnote{The energies of those two states differ by not more than 0.1 cm$^{-1}$ in the spin-orbit
ROHF, CASSCF, and MRCI calculations.
}.
Given the tetragonal distortions in Ba$_2$IrO$_4$, with out-of-plane ($z$-axis) Ir-O bonds significantly
stretched as compared to the in-plane ($x$/$y$) bonds \cite{Ba214_structure}, this degeneracy is somewhat surprising.
Using Eqs.~(\ref{a4}), this means that two of the diagonal couplings of $\bs{\Gamma}$ are equal,
$\Gamma_{zz}=\Gamma_\perp$, which further implies $\Gamma_{\parallel}\!=\!-2\Gamma_\perp$.
The interaction terms in (\ref{a2}) and (\ref{a3}) can then be rewritten as
\bea\label{a5}
&&\mc{H}_{\langle ij\rangle \parallel x} = \bar J\,{\tilde {\bf S}}_i\!\cdot\! {\tilde {\bf S}}_j + \bar \Gamma_{\parallel}\,{\tilde S}_i^{x} {\tilde S}_j^{x}~,\nonumber\\
&&\mc{H}_{\langle ij\rangle \parallel y} = \bar J\,{\tilde {\bf S}}_i\!\cdot\! {\tilde {\bf S}}_j + \bar \Gamma_{\parallel}\,{\tilde S}_i^{y} {\tilde S}_j^{y}~,
\eea
where $\bar{J} \!\equiv\! J \!+\! \Gamma_{\perp}$ and 
$\bar{\Gamma}_{\parallel}\!\equiv\!-3\Gamma_{\perp}$.
Quantum chemistry results for $\bar J$ and $\bar \Gamma_{\parallel}$ are provided on the lowest
line in Table\,\ref{table_Js}.

The value computed for the Heisenberg $\bar J$ within the ROHF approximation, $-12$ meV (see Table
\ref{table_Js}), is sizable and close to the results computed in square-lattice $3d^9$ Cu oxides
(see, e.g., Ref.~\onlinecite{NOCI_J_oosten96}).
It accounts only for direct exchange, since no (intersite) excitations are allowed.
In contrast to the ROHF $\bar J$, the anisotropic $\bar \Gamma_{\parallel}$ is AF by ROHF.

\begin{table}[!t]
\caption{
Energy splittings for the four lowest spin-orbit states of two NN IrO$_6$ octahedra
and the corresponding effective coupling constants in Ba214, at different levels of 
approximation (all in meV).
} \label{table_Js}
\begin{ruledtabular}
\begin{tabular}{lccc}
States/Method                                                                  &HF+SOC &CAS+SOC &CI+SOC \\
\hline
$\Psi_{\mathrm{S}}(A_{1g})= (\uparrow\downarrow -\downarrow\uparrow)/\sqrt2$   &$12.2$ &$0.0$   &$0.0$  \\
$\Psi_{\mathrm{3}}(A_{1u})= (\uparrow\uparrow   -\downarrow\downarrow)/\sqrt2$ &$0.0$  &$37.5$  &$65.0$ \\
$\Psi_{\mathrm{1}}(B_{2u})= (\uparrow\downarrow +\downarrow\uparrow)/\sqrt2$   &$0.2$  &$38.2$  &$66.7$ \\
$\Psi_{\mathrm{2}}(B_{1u})= (\uparrow\uparrow   +\downarrow\downarrow)/\sqrt2$ &$0.2$  &$38.2$  &$66.7$ \\
($\bar J$, $\bar \Gamma_{\parallel}$)                                          &($-12.0$,$0.4$)
                                                                               &($ 37.5$,$1.4$)
                                                                               &($ 65.0$,$3.4$) \\
\end{tabular}
\end{ruledtabular}
\end{table}

With correlated wave functions, CASSCF and MRCI, the singlet $\Psi_{\mathrm{S}}$ becomes the
ground state, well below the triplet components $\Psi_{\mathrm{1}}$, $\Psi_{\mathrm{2}}$, and $\Psi_{\mathrm{3}}$.
This shows that the largest energy scale is here defined by the isotropic Heisenberg exchange $\bar J$
($\bar J\!>\!0$).
In the CASSCF approximation, only intersite $d$--$d$ excitations {\` a} la Anderson \cite{Anderson_1950}
are accounted for, i.e., polar $t_{2g}^6$--\,$t_{2g}^4$ configurations.
Again, the CAS+SOC $\bar J$, 37.5 meV, is very similar to the CASSCF $J$'s in layered $3d^9$ cuprates
\cite{Cu_d9_J_goddard_88,Cu_d9_J_martin_88,NOCI_J_oosten96}.
It is seen in Table~\ref{table_Js} that the configuration-interaction treatment, which now includes as
well $t_{2g}^5e_g^1$--\,$t_{2g}^4$ and O $2p$ to Ir $5d$ charge-transfer virtual states, enhances $\bar J$
by about 70\% as compared to the CAS+SOC value, somewhat less spectacular than the ratio between
the configuration-interaction and CASSCF $J$'s in layered cuprates.
In the latter compounds, this ratio is 3 to 4~\cite{NOCI_J_oosten96,CuO_J_Illas_00}.

If we include in the MRCI treatment only the six Ir $t_{2g}$ orbitals, $\bar J$ is 49.1 meV (not
shown in Table~\ref{table_Js}).
The difference between the latter number and the CAS+SOC value given in Table~\ref{table_Js} is indicative 
of the role of excitation processes via the Ir $5d$ $e_g$ levels.
The further increase from 49.1 to 65 meV is due to excitations that additionaly involve the
bridging O $2p$ orbitals.
The data in Table~\ref{table_Js} also show that the correlation treatment very much enlarges the symmetric
anisotropic coupling $\bar \Gamma_{\parallel}$, from 0.4 by ROHF to 3.4 meV by MRCI.

\section{Comparison to effective superexchange models}\label{sec:IV}

For the Mott-like insulating regime occurring in the iridates \cite{IrO_mott_kim_08,214Ir_kim_2009,IrO_kitaev_jackeli_09},
an effective superexchange model can be in a first approximation set up by considering the
leading excited configurations with two holes at the same Ir site.
With corner-sharing octahedra and straight Ir-O-Ir bonds along the $x$ axis, the intersite
$d$--$d$  hopping takes place via both in-plane $p_y$ and out-of-plane $p_z$ $\pi$-type O
orbitals.
The relevant effective hopping integrals are
$t_1=(t_{pd}^{\pi})^2/|\epsilon^{xy}_d-\epsilon^{y}_p|$ for the in-plane, $xy$ pair of NN Ir $t_{2g}$
functions and
$t_2=(t_{pd}^{\pi})^2/|\epsilon^{xz}_d-\epsilon^{z}_p|$ for the out-of-plane, $xz$ $t_{2g}$ functions.
$\epsilon^{y/z}_p$ and $\epsilon^{xy/xz}_d\!=\epsilon_{1/2}$ are here crystal-field split
energy levels while the $p$--$d$ $\pi$-type hopping amplitude $t_{pd}^{\pi}$ is assumed to be the
same for both chanels.

For tetragonal distortions, $\epsilon_{1}\neq\epsilon_{2}$, $\epsilon^{y}_p\neq\epsilon^{z}_p$
and therefore $t_1$ and $t_2$ may acquire quite different values.
A hole hopping between NN Ir ions is then described by the Hamiltonian
\bea\label{a6}
H^{ij}_{\mathrm{hop}}= \sum\limits_{m=1,2}\sum\limits_{\sigma=\uparrow,\downarrow} \left(t_m
d^{\dagger}_{im\sigma}d_{jm\sigma} + h.c.\right)\,, 
\eea
where $d^{\dagger}_{im\sigma} (d_{im\sigma})$ is the creation (annihilation) operator of a hole
with spin $\sigma$ in the orbital $d_{xy}$ for $m=1$ and $d_{xz}$ for $m=2$ at site $i$.
For a bond along the $y$ axis, $p_y$ is replaced by $p_x$, $d_{xz}$ by $d_{yz}$,
$\epsilon^{y}_p\!=\epsilon^{x}_p$, $\epsilon_3\!=\epsilon^{yz}_d\!=\epsilon^{xz}_d\!=\epsilon_2$,
and the hopping Hamiltonian in (\ref{a6}) has the same form.

The interaction of two holes in the $t_{2g}$ subshell is described by Hund's coupling $J_H$
and the Coulomb repulsion integrals $U_{mm'}\simeq U - 2J_H$, if $m\neq m'$, and $U_{mm} = U$.
While the isotropic exchange is related to second-order processes that concern transitions
between the lowest spin-orbit Kramers doublets, i.e., $J\sim t_{1/2}^2/U$,
the symmetric anisotropy is entirely determined by third-order processes that involve excited
Kramers doublets, i.e., is dependent on $t_{1/2}^2J_H/U^2$.

The lowest Kramers doublet wave functions
\bea\label{a7}
|\tilde {\uparrow}\rangle &=& \sin\theta|xy,\uparrow\rangle +
\frac{\cos\theta}{\sqrt{2}}\left(i|xz,\downarrow\rangle + |yz,\downarrow\rangle\right) \nonumber \\ 
|\tilde {\downarrow}\rangle &=& \sin\theta|xy,\downarrow\rangle -
\frac{\cos\theta}{\sqrt{2}}\left(i|xz,\uparrow\rangle - |yz,\uparrow\rangle\right) 
\eea
as well as those for the excited Kramers doublets are here parametrized as in 
Ref.~[\onlinecite{IrO_kitaev_jackeli_09}], with the angle $\theta$ given by
$\tan(2\theta)=2\sqrt{2}\lambda/(\lambda-2\Delta)$ while $\Delta=\epsilon_2^d-\epsilon_1^d$
is the tetragonal $t_{2g}$ splitting.

By collecting the second- and third-order processes in this effective superexchange model,
we arrive at the pseudospin Hamiltonian in (\ref{a2}), with
\bea\label{a8}
J\!&=&\!\frac{4}{U}\left(t_1\sin^2\theta + \frac{t_2}{2}\cos^2\theta\right)^2  + \gamma\,,\nonumber\\
\Gamma_{\parallel}\!&=&\! - \eta\frac{3(t_1 - t_2)^2}{U}\sin^2\theta\cos^2\theta - \gamma\,,\nonumber\\
\Gamma_{\perp}\!&=&\! - \eta\frac{3t_1^2}{U}\sin^2\theta\cos^2\theta -\gamma\,, \nonumber\\
\Gamma_{zz} \!&=&\! - \eta\frac{3t_2^2}{2U}\cos^4\theta -\gamma\,.  
\eea
Here $\eta=J_H/U$ and $\gamma=-\frac{\eta}{U}\cos^2\theta [(t_1-t_2)^2\sin^2\theta+t_1^2\sin^2\theta+\frac{1}{2}t_2^2\cos^2\theta]$.

Now, for $\Gamma_{zz}=\Gamma_{\perp}$, the model described by (\ref{a5}) displays uniaxial compass-like anisotropy~\cite{KK1973}.
That is obviously the case for perfect, cubic octahedra with $\Delta\!=\!0$, $t_1\!=\!t_2\!=\!t$, and $\cos\theta_c\!=\!\sqrt{2}\sin\theta_c\!=\!\sqrt{2/3}$. 
In the cubic limit we further have from Eqs.~(\ref{a8}):
$J^c\!=\!(16/9)t^2/U+\gamma^c$,
$\gamma^c=-(4\eta/9)t^2/U$,
$\Gamma_{\parallel}^c\!=\!-\gamma^c$, and
$\Gamma_{\perp}^c\!=\!\Gamma_{zz}^c\!=\!(-2\eta/3)t^2/U -\gamma^c$.

For tetragonal distortions as found in Ba$_2$IrO$_4$ \cite{Ba214_structure},
$\Gamma_{\perp}\!=\!\Gamma_{zz}$ implies that $(t_2/t_1)^2\!=\!2\tan^2\theta$.
As measure of how large the departure from the cubic limit is we can take the ratio
between the tetragonal $t_{2g}$ splitting $\Delta$ and the strength of the spin-orbit coupling $\lambda$.
The quantum chemistry calculations yield $\Delta\!=\!65$ meV in Ba$_2$IrO$_4$, see the
discussion in Ref.~[\onlinecite{RhIr214_vmk_13}], in agreement with estimates based on experiment
\cite{Ba214_rixs_moretti_2014}.
A direct estimate of the spin-orbit coupling can be also obtained from the splitting of the
$j\!=\!1/2$ and $j\!=\!3/2$ $t_{2g}^5$ states for idealized cubic octahedra.
It turns out that for perfect octahedra $\lambda\!=\!0.47$ eV \cite{113Ir_bogdanov_2012,CaIrO3_moretti_14,RhIr214_vmk_13},
close to values of 0.4--0.5 eV earlier derived from electron spin resonance and optical measurements
on $5d^5$ ions \cite{SOC_d5_dingle65,SOC_d5_allen72,SOC_d5_andlauer76,SOC_d5_ping09}.
The ratio $\Delta/\lambda$ is therefore rather small, $\approx$0.15.

Estimates for the parameters that enter the effective superexchange model can be most easily obtained in the cubic limit.  
Using Eqs.~(\ref{a8}) we find that $\bar \Gamma_{\parallel}/\bar J\!\approx\!(3/8)\eta$. 
The CI+SOC values of Table~\ref{table_Js}, $\bar \Gamma_{\parallel}\!=3.4$ and $\bar J\!=65$ meV, then
lead to $\eta\!\approx\!0.14$ and $4t^2/U\!\approx\!149$ meV.
Interestingly, estimates of the hopping integral $t$ from calculations based on density-functional
theory (DFT) are $t_{\mathrm{DFT}}\!\approx\!260$ meV, while the on-site Coulomb repulsion comes out from
constrained calculations in the random phase approximation (RPA) as $U_{\mathrm{RPA}}\!\approx\!1.65$
eV \cite{IrO_mott_arita_12}.
The ratio $4t_{\mathrm{DFT}}^2/U_{\mathrm{RPA}}$ is therefore $\approx$164 meV, close to the result
derived on the basis of the CI+SOC effective couplings listed in Table\,I.
On the other hand, the $\eta$ parameter extracted from the periodic DFT calculations
\cite{IrO_mott_arita_12} is $\eta_{\mathrm{DFT}}\!\approx\!0.08$, much smaller than the above value of 0.14.
Using the latter value for $\eta$, $\eta_{\mathrm{DFT}}\!\approx\!0.08$, an estimate for the symmetric anisotropic coupling
$\bar \Gamma_{\parallel} = \frac{3}{8}\eta \bar J$ would be significantly smaller than the
quantum chemistry result.

\section{Ground state phase diagram}\label{sec:V}

Having established the strength of the dominant in-plane exchange interactions and anisotropies,
we now turn to the nature of the magnetic ground state of Ba$_2$IrO$_4$, focusing first on a single
square-lattice IrO$_2$ layer.
In the classical limit, the compass-Heisenberg model defined by Eqs.~(\ref{a5}) has an accidental SO(2) ground-state degeneracy,
with spins pointing along any direction in the basal $xy$-plane~\cite{KK1973,Zohar2005,Fabien2012}.
This degeneracy is eventually lifted via thermal~\cite{Mishra2004,Zohar2004,Sandro2010} or
quantum~\cite{Aharony98,Khaliullin2001,Dorier2005,Sandro2010} order-by-disorder effects,
whereby harmonic spin wave fluctuations select the states with spins pointing either along the $x$ or $y$ axis. 
This is however in sharp contrast to experiments which below $\sim$240 K show basal-plane AF order
with magnetic moments along the [110] direction~\cite{BoseggiaPRL_13}.
It
indicates additional anisotropies in the system, large enough to overcome the energy gain from
the order-by-disorder mechanism.

\begin{figure}[b!]
\includegraphics[angle=0,width=0.90\columnwidth]{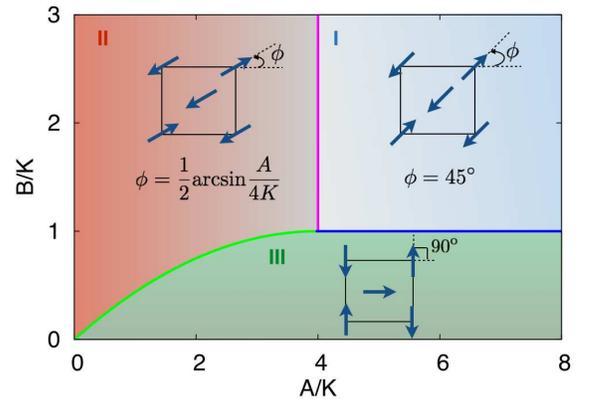}
\caption{Ground-state phase diagram of the model described in Sec.~\ref{sec:V},
including the single-layer Heisenberg-compass terms of (\ref{a5})
plus the effect of interlayer couplings, see text.}
\label{phase_diagram}
\end{figure}

The situation is actually analogous to several $3d^9$ Cu oxides with the same ``214'' crystal structure as Ba$_2$IrO$_4$.
It has been shown that in cuprates that particular type of AF order is selected by a subtle 
interplay between in-plane and interlayer interactions, as discussed in detail in Ref.~[\onlinecite{Aharony98}].
Assuming that qualitatively the same 3D mechanism is applicable to Ba$_2$IrO$_4$, we analyze below
the main contributions to the expression of the 3D ground-state energy and derive a generic phase diagram.
This exercise provides useful insights into the dependence of the ground-state spin configuration on
various interaction parameters in 214 iridates.

It turns out that the most important effects competing with the in-plane NN interactions concern
(i) the frustrating nature of the isotropic interlayer exchange and (ii) the symmetric part of the
anisotropic exchange between layers.
To show this we proceed by parametrizing the global spin direction in each basal plane by an angle
$\phi_n$, where $n$ is the layer index,  and by writing down all relevant energy contributions.

The first contribution is the zero-point energy  (per spin) coming from the order-by-disorder mechanism in each individual layer,
$E_{\rm ZP,2D}(\{\phi_n\})=\sum_n \mc{E}_{\rm ZP,2D}(\phi_n)$, where
\be
\mc{E}_{\rm ZP,2D}(\phi)=\frac{1}{2N}\sum\limits_{\bf q}\left( \omega_{+}({\bf q}) + \omega_{-}({\bf q}) \right)
\label{a9}
\ee
and $\omega_\pm(\vec{q})$ are the two spin wave branches,
for which explicit expressions are provided in Appendix \ref{app1}.
A numerical analysis of Eq.~(\ref{a9}), using the {\it ab initio} quantum chemistry values for
the in-plane NN effective couplings (see Sec.~\ref{sec:III}), shows that $\mc{E}_{\rm ZP,2D}(\phi)$ is almost identical to the expression
\begin{eqnarray}
\mc{E}_{\rm ZP,2D}(\phi)= -K\cos(4\phi) + E_0, \label{a10}
\end{eqnarray}
with $K\!=\!0.86$ $\mu$eV and $E_0\!=\!56.55$ meV.

We now turn to the second contribution to the energy, which stems from the interlayer isotropic
exchange $J_{\sf out}$.
Despite being the dominant portion of the interlayer interactions, its total contribution to the
energy vanishes in the mean-field sense due to geometric frustration in the 214 structure, see
Fig.~\ref{figBa214}. 
Yet quantum fluctuations driven by $J_{\sf out}$ still give rise to a zero-point energy contribution
\be
E_{\rm ZP,3D}(\{\phi_n\})= -B\sum\limits_{n}\cos(2\phi_n - 2\phi_{n+1})\,, \label{a11}
\ee
where $B\simeq 0.032 J_{\sf out}^2/(2J_{\sf av})$ and
$J_{\sf av}=J+(\Gamma_\parallel+\Gamma_\perp)/2$~~\cite{Yildirim96}.
Since $B$ is positive for any sign of $J_{\sf out}$, this contribution favors collinearity of the staggered magnetization in adjacent layers.

The third contribution to the energy comes from the anisotropic portion of the interlayer couplings.
We first note that the antisymmetric DM component vanishes by symmetry since the midpoint of
each of these out-of-plane NN Ir-Ir links is an inversion center.
The remaining, symmetric portion, can be described by a traceless second-rank tensor $\bs{\Gamma}_{\sf out}$.
The structure of the latter is simplified by using the fact that the out-of-plane NN Ir-Ir links are $C_2$
axes, additionally perpendicular to reflection planes.
Adding up the four tensors (related to each other by symmetry) from all four NN bonds above/below
the reference layer gives \cite{Aharony98}
\begin{eqnarray}
E_{\rm aniso,3D}= -A\sum\limits_{n}\sin(\phi_n + \phi_{n+1}), \label{a12}
\end{eqnarray}
where the constant $A$ is fixed by the elements of $\bs{\Gamma}_{\sf out}$.

The total energy now reads
\be
E = E_{\rm ZP,2D} + E_{\rm ZP,3D} + E_{\rm aniso,3D}~.
\ee
It can be minimized analytically as described in Appendix~\ref{app2} by
working it out for a
bilayer of Ba214.
The resulting phase diagram in the $(A/K,B/K)$ plane is shown in Fig.~\ref{phase_diagram} for positive $A$ (the phase diagram for $A\!<\!0$ is identical, see Appendix~\ref{app2}),
and hosts three different phases, two collinear (phases I and II) and one noncollinear (phase III).

In phase I, the staggered magnetizations point along one of the $\langle110\rangle$ axes and
the relative directions between adjacent planes are regularly collinear or anticollinear.
In phase III, the AF magnetization prefers one of the $\langle100\rangle$ axes and the relative
directions in adjacent planes are now perpendicular to each other.
Finally, in phase II, the relative directions between adjacent planes are again either collinear
or anticollinear but the staggered magnetizations in each layer rotate in the basal
plane as a function of $A/K$, see Appendix~\ref{app2}.
Importantly, the degeneracy is not completely lifted by the above couplings.
As explained in Appendix~\ref{app2}, all phases have an Ising degree of freedom per layer,
which comes from the fact that the energy remains the same if we flip all spins within a
given layer.
This remaining macroscopic degeneracy may eventually be lifted via higher-order processes
or farther-neighbor couplings, see for example the discussion in [\onlinecite{Yildirim96}].
The collinear AF structure observed experimentally~\cite{BoseggiaPRL_13} in Ba$_2$IrO$_4$ can now
be naturally explained provided that $A$ and $B$ fall into the broad region of phase I in the phase
diagram of Fig.~\ref{phase_diagram} and by taking into account
removal of the macroscopic Ising degeneracy by the mechanism mentioned above.

As pointed out by Boseggia {\it et al.}\cite{BoseggiaPRL_13}, the AF component of the ordered momenta
in the 214 iridates Sr$_2$IrO$_4$ and Ba$_2$IrO$_4$ is essentially identical -- 
in Sr$_2$IrO$_4$, the canted AF state is characterized by an AF vector aligned along the $\langle110\rangle$
direction and a residual FM moment confined to the same basal plane.
Staggered rotation of the IrO$_6$ octahedra as realized in Sr$_2$IrO$_4$ requires 
the more general single-layer Hamiltonian of Eq.~(\ref{a1}) \cite{IrO_kitaev_jackeli_09,Ir214_perkins_13}, 
with a DM vector along the $z$ axis and a biaxial easy-plane symmetric anisotropy described in our
notation by two independent diagonal components $\Gamma_{\parallel}>0$ and $\Gamma_{zz}>0$.
This model correctly explains the canting angle of the basal-plane AF order \cite{Ir214_perkins_13}
but fails in predicting the AF vector alignment along one of the $\langle110\rangle$ axes.
The reason is that the two additional anisotropies, ${\bf D}||z$ and $\Gamma_{zz}>0$, do not
remove the SO(2) basal-plane ground-state degeneracy, at least not in the classical limit.
This accidental degeneracy can however again be lifted via the 3D mechanism discussed above, to 
arrive to an AF ordering pattern similar to that of Ba$_2$IrO$_4$ \cite{BoseggiaPRL_13}.

\section{Conclusions}

While {\it ab initio} quantum chemistry techniques have been earlier used to derive the sign 
and strength of the symmetric anisotropic (Kitaev) interactions in $5d^5$ iridates
with edge-sharing IrO$_6$ octahedra \cite{Ir213_katukuri_13}, we here employ the same
methodology to clarify the signs and magnitude of the symmetric anisotropic couplings for
corner-sharing octahedra in the square-lattice compound Ba$_2$IrO$_4$.
The {\it ab initio} results reveal effective uniaxial anisotropy, although the actual symmetry of 
each of the in-plane Ir-Ir links is lower than $D_{4h}$.
The anisotropic effective coupling constants are as large as 3.5 meV, comparable in strength with 
the anisotropic Kitaev exchange in honeycomb Na$_2$IrO$_3$ \cite{Ir213_katukuri_13}.
However, in contrast to Na$_2$IrO$_3$, the largest energy scale is here defined by the Heisenberg
$J$, with $J\!\approx\!65$  meV.
The latter value agrees with estimates based on resonant inelastic x-ray scattering 
measurements on 214 iridates \cite{Kim12}.
Given the uniaxial structure of the exchange coupling tensor, the relevant in-plane (pseudo)spin
model is a Heisenberg-compass type of model.
Yet to understand the experimentally determined AF ordering pattern, with spins along the [110] direction
\cite{BoseggiaPRL_13}, interlayer interactions must be included in the effective Hamiltonian.
Further investigations are now carried out in our group for quantifying the strength of
Dzyaloshinskii-Moriya couplings for the closely related 214 compound Sr$_2$IrO$_4$, displaying
bent Ir-O-Ir links.
Another interesting issue is the dependence of the in-plane anisotropic couplings, their signs
in particular, on pressure \cite{Sr2IrO4_xmcd_2012} and strain \cite{Ir214_lupascu_13}, in
both Ba$_2$IrO$_4$ and Sr$_2$IrO$_4$.

\section{Acknowledgements}
L.~H. acknowledges financial support from the German Research Foundation (Deutsche Forschungsgemeinschaft, DFG).

\appendix
\begin{widetext}

\section{Spin wave dispersions}\label{app1}
In the magnetic Brillouin zone, where $\sum_{\vec{q}}=N/2$, there are two spin wave branches,
with dispersions
given by~\cite{Aharony98}
\be
\omega_{\pm}(\vec{q})=4J_{\sf av} S \sqrt{(1\mp B_{\vec{q}})^2+A_{\vec{q}}^2}\,.
\ee
In this expression,
$S=1/2$, $J_{\sf av}=J+(\Gamma_\parallel+\Gamma_\perp)/2$,
\be
A_{\vec{q}} = \frac{1}{4J_{\sf av}} \left[ J_1 \cos(q_x a)+J_2 \cos(q_y a)\right],~~
B_{\vec{q}} = -\frac{1}{4J_{\sf av}} \left[ J_3 \cos(q_x a)+J_4 \cos(q_y a)\right]\,,
\ee
and
\bea
&&J_1=2J+\Gamma_{zz}+\Gamma_\parallel\sin^2\phi + \Gamma_\perp \cos^2\phi ,~~~
J_2=2J+\Gamma_{zz}+\Gamma_\parallel\cos^2\phi + \Gamma_\perp \sin^2\phi \nonumber\\
&&J_3= -\Gamma_{zz}+\Gamma_\parallel\sin^2\phi +\Gamma_\perp\cos^2\phi ,~~~
J_4=-\Gamma_{zz} +\Gamma_\parallel\cos^2\phi +\Gamma_\perp\sin^2\phi~.
\eea

These can be rewritten in terms of the coupling constants $\bar J$ and $\bar \Gamma_{\parallel}$ entering the Hamiltonian 
terms in (\ref{a5}) by making the replacements 
$J=\bar J+\frac{1}{3}\bar \Gamma_{\parallel}$, $\Gamma_{\parallel}=\frac{2}{3}\bar \Gamma_{\parallel}$,
and $\Gamma_{\perp}=\Gamma_{zz}=-\frac{1}{3}\bar \Gamma_{\parallel}$.

\section{Energy minimization for a bilayer}\label{app2}

The ground-state magnetic energy of the layered system  can be
written as a sum over bilayer contributions (per spin and per
layer): \bea
E(\phi_1,\phi_2)=-\frac{K}{2}[\cos(4\phi_1)+\cos(4\phi_2)]
-B\cos[2(\phi_1-\phi_2)] -A\sin(\phi_1+\phi_2) \nonumber \\
= -K \cos(2\phi_+) \cos(2\phi_-) -B\cos(2\phi_-)-A\sin\phi_+,
\nonumber \eea
where the angles $\phi_1$ and $\phi_2$ define orientations
(say, with respect to the $x$ axis) in two adjacent
planes and $\phi_\pm=\phi_1\pm\phi_2$.
We note that both $K$
and $B$ are positive. In the subsequent discussion, the coupling
$A$ is chosen positive as well by taking into account the fact that for
$A<0$ the simultaneous change of signs, $\phi_1\to -\phi_1$ and
$\phi_2\to -\phi_2$, retains the expression for $E(\phi_1,\phi_2)$
invariant.

Minimizing $E(\phi_1,\phi_2)$  we find  four possible extrema
solutions for $\phi_1$ and $\phi_2$ and the respective energies ($n$
and $m$ are integers): \bea \phi_-^{(1)}=m\pi, \quad
\phi_+^{(1)}=\frac{\pi}{2}+2n\pi ,\quad
 E^{(1)}= K -B - A, \label{b1} \eea which is possible if $B>K$;
 \bea
\phi_-^{(2)}=m\pi, \quad \phi_+^{(2)}=\arcsin\frac{A}{4K}+2n\pi,
\quad E^{(2)}= -K -B - \frac{A^2}{8K}, \label{b2} \eea with the
requirement $A<4K$; \bea \phi_-^{(3)}=(2m+1)\frac{\pi}{2}, \quad
\phi_+^{(3)}=\frac{\pi}{2}+2n\pi , \quad E^{(3)}= B -K - A,
\label{b3} \eea
 which is possible if $B<K$;
\begin{eqnarray}
\sin\phi_+^{(4)}=\sqrt{\frac{1+B/K}{2}}\ , \quad
\cos(2\phi_-^{(4)})=\frac{A}{4K}\sqrt{\frac{2}{1+B/K}}\ ,\quad
E^{(4)} = - A \sqrt{\frac{1+B/K}{2}}\ , \label{b4}
\end{eqnarray}
 which may occur in
the parameter region $B\!<\!K$, $A\!<\!K\sqrt{\frac{1+B/K}{2}}$.

Comparison of the energies of the four possible
ground-state configurations shows that three of them, from
(\ref{b1}) to (\ref{b3}), occur in different domains of the $A$--$B$
parameter space.
In the region $B>K$ and $A>4K$, the most
stable is the configuration (\ref{b1}) with
$\phi_1^{(1)}=\frac{\pi}{4}+n\frac{\pi}{2}$ and
$\phi_2^{(1)}=\phi_1^{(1)}-m\pi$, which means that the spins
(staggered magnetizations) are along one of $\langle110\rangle$ axes and in two
adjacent planes the spin alignment is either collinear or anticollinear.
Next, in the region with $B\!>\!K$, $0\!<\!A\!<\!4K$, the second configuration
(\ref{b2}) with
$\phi_1^{(2)}=\frac{1}{2}\arcsin(A/4K)+n\frac{\pi}{2}$ and
$\phi_2^{(1)}=\phi_1^{(1)}-m\pi$ is realized. Here, the
collinear/anticollinear alignment in successive layers  still
persists.
However, the preferred direction is specified by $A/4K$.
In the region with $B\!<\!K$, $A\!>\!4K$, the third configuration (\ref{b3})
with $\phi_1^{(3)}= m\frac{\pi}{2}$ and
$\phi_2^{(3)}=\phi_1^{(3)}- \frac{\pi}{2}-m\pi$ is the most stable, which corresponds
to having the magnetization  along one of the $\langle100\rangle$ axes with two
directions in successive layers being rotated by $90^{\circ}$.
Finally, for $B\!<\!K$ and $A\!<\!4K$, the fourth solution (\ref{b4}) has
the highest energy and two of the other configurations, i.e., (\ref{b2}) and
(\ref{b3}), compete to give the phase boundary depicted in Fig.~\ref{phase_diagram}.

{\it Ising degrees of freedom} ---
It is clear that the above classical minima of a Ba214 bilayer are also the minima of
the infinite system. In all phases, however, there is still an Ising degree of freedom per layer,
which is not fixed by the couplings considered here. In phase I for example,
we may flip the directions of all spins in any plane, since the energy is the
same for both collinear and anticollinear relative orientations between adjacent planes.
The eventual removal of this remaining macroscopic Ising degree of freedom must originate from higher-order 
processes or farther-neighbor couplings~\cite{Yildirim96}.

\end{widetext}


%
\end{document}